\documentclass[aps,prc,preprint,showpacs,amsmath,amssymb,nofootinbib]{revtex4}
\usepackage {lscape}
\usepackage{graphicx}
\usepackage{amsmath}

\begin{document}

%\title[High-spin structures of Te isotopes]{Shell model results for high-spin structures of $^{124-127}$Te isotopes : Competition of proton- and 
%neutron-pair breakings}
%\title{Shell model study for high-spin structures of  the near-spherical nuclei $^{91,92}$Zr}
\title{High-spin structures of  the near-spherical nuclei $^{91,92}$Zr}
%\author{ A and B}
\author{P.C. Srivastava\footnote{Email: pcsrifph@iitr.ac.in}}
\address{Department of Physics, Indian Institute of Technology,
  Roorkee - 247667, India}

\date{\hfill \today}
\begin{abstract}% 
In the present work, we have interpreted recently available experimental data  for
high-spin states of the near-spherical nuclei  $^{91,92}$Zr, using the shell-model calculations within the
full $f_{5/2}$, $p_{3/2}$, $p_{1/2}$,  $g_{9/2}$ model space for protons and valence neutrons in $g_{9/2}$, $g_{7/2}$, $d_{5/2}$ orbits.
We have employed a truncation for the neutrons due to huge matrix dimensions, by allowing one neutron excitation from
$g_{9/2}$ orbital to $d_{5/2}$ and $g_{7/2}$ orbitals. Results are in good agreement with the available experimental data.
Thus, theoretically, we have identified the structure of many high-spin
states, which were tentatively assigned in the recent experimental work. The $^{91}$Zr $21/2^+$ isomer 
lies at low-energy region due to fully aligned spins of two $g_{9/2}$ protons and one $d_{5/2}$ neutron.
\end{abstract}


\maketitle

\section{Introduction}
\label{s_intro}
The nuclei around $N=50$ region have recently attracted considerable experimental and theoretical attention
\cite{ref1,ref2,ref3,ref4,ref5,ref6,9192Zr,honma,yoshi,nakada,gepcs,mpla,62ga,1a,88Zr,89Zr,89Nb,91Nb,pittel,patt,pajtler,bingham,pcappb}. The excitation of one neutron from 
$g_{9/2}$ to $d_{5/2}$ orbital is responsible for high-spin level structure of As,  Sr, Y, Zr, Mo, Tc, Ru and Rh nuclei, with $N$ $\sim$ 50 \cite{1a,1b,1c,2,3,4,5,6,7,8,9,10}. 
The importance of monopole fitted effective interaction, inclusion of higher orbitals across $N=50$ shell
and seniority quantum number are discussed in  Refs. \cite{bansal64,otsu1,otsu2,prog3,talmi}.

Recently, high-spin level structure of the semi-magic nucleus $^{91}$Nb has been
investigated via $^{82}$Se($^{14}$N,$5n$)$^{91}$Nb reaction \cite{91Nb}.
The importance of neutron particle-hole excitation across $N=50$ shell gap
is reported to describe the states above $\sim$ $(21/2) \hbar$. The high-spin
level structures of near-spherical nuclei $^{91,92}$Zr have been reported
in \cite{9192Zr}. As we move from  $^{91}$Zr to $^{92}$Zr, a possible 
reduction in the energy gap between $p_{3/2}$ and $p_{1/2}$ orbitals are indicated, 
since one more neutron is excited across the $N=50$ gap.
For $^{92}$Zr, Werner  {\it et al.}, reported the pure neutron configuration to the $2_1^+$ state,
while the second excited quadrupole state shows the signatures of the one-phonon mixed-symmetric $2^+$ state \cite{wernerplb}.
The results of an in-beam study of high-spin structure of $^{94,95}$Mo at
GASP, Legnaro was reported in \cite{9495Mo}. In this work the authors also reported shell model results
in  $\pi (p_{1/2},g_{9/2},d_{5/2})$ and $\nu (p_{1/2}, g_{9/2}, g_{7/2}, d_{5/2}, d_{3/2}, s_{1/2})$
space.  With this space the yrast states $12_1^+$ and $15_1^-$  in $^{94}$Mo and those up to $25/2_1^+$ and $31/2_1^-$ in $^{95}$Mo are explained.
The importance of possible contribution from the $Z = 38$, $N = 50$ core, especially that of neutron excitations
from the $g_{9/2}$ orbital has been reported in this work.  

In the present paper, we report a systematic study of shell-model results for $^{91,92}$Zr. The aim of present work is to explain the structure of recently available high-spin experimental data for these two isotopes \cite{9192Zr}.

The work is organized as follows: details of the calculation and effective interaction are given in section II, 
the shell model results and a comprehensive
comparison with experimental data and wave-function analysis
are given in section III. Finally, concluding remarks are
drawn in section IV.

\section{Details about effective interaction}\label{discuss_hm}
In this work, the results of recently available experimental data of $^{91,92}$Zr have been interpreted
with large scale shell-model calculations. The model space consists of 
full $f_{5/2}$, $p_{3/2}$, $p_{1/2}$, $g_{9/2}$ orbitals for protons and valence neutrons in $p_{1/2}$, $g_{9/2}$,
$g_{7/2}$, $d_{5/2}$ orbitals. Since
the dimension of matrices are very large, we allowed neutrons to occupy only 
$g_{9/2},d_{5/2},g_{7/2}$ orbitals.
In the present work, we have taken $^{68}$Ni as a core.
We have  allowed one neutron excitation from
$g_{9/2}$ orbital to $d_{5/2}$ and $g_{7/2}$ orbitals.
The calculations have been carried out
with GWBXG effective interaction. The single-particle energies (in MeV) for protons orbital
are $f_{5/2}$= -5.322, $p_{3/2}$= -6.144, $p_{1/2}$= -3.941, $g_{9/2}$ = 
-1.250, and for neutrons orbital are $g_{9/2}$ = -2.597, $g_{7/2}$ = +5.159,
$d_{5/2}$= +1.830. 
We have performed shell model calculation with shell model code NuShellX \cite{nushell}.
The GWBXG effective interaction is constructed with different interactions. 
The original 974 two-body matrix elements (TBMEs) were taken from
bare G-matrix of H7B potential \cite{int1}. 
The bare G-matrix is not reasonable because of the space
truncation and the interaction should be renormalized by taking into account the core-polarization effects.
Here, the present G-matrix effective interaction is further tuned by modifying matrix elements with fitted interactions
by following way. The 65 TBMEs for proton orbitals 
were replaced with the effective values of
Ji and Wildenthal \cite{int2}. The TBMEs connecting the $\pi(p_{1/2},
g_{9/2)}$  and the $\nu(d_{5/2}, s_{1/2})$ orbitals were replaced by those from
the work of Gloeckner \cite{int3}, and those between the $\pi(p_{1/2},
g_{9/2})$ and the $\nu(p_{1/2}, g_{9/2})$ orbitals were replaced with
those of Serduke {\it et al.} \cite{int4}. Previously,
shell model results with this interaction for the positive-parity yrast states of the neutron-rich
$^{89}$Rb, $^{92}$Y, and $^{93}$Y nuclei were reported in Ref. \cite{gwbxg_cite}.

%%%%%%%%%%%%%%%%%%%%%%%%%%%%%%%%%%%%%%%%%%%%%%%%%%%%%%%  
%%%%%%%%%%%%%%%%%%%%%%%%%%%%%%%%%%%%%%%%%%%%%%%%%%%%%%%%%%
\begin{figure}
\includegraphics[width=17.0cm]{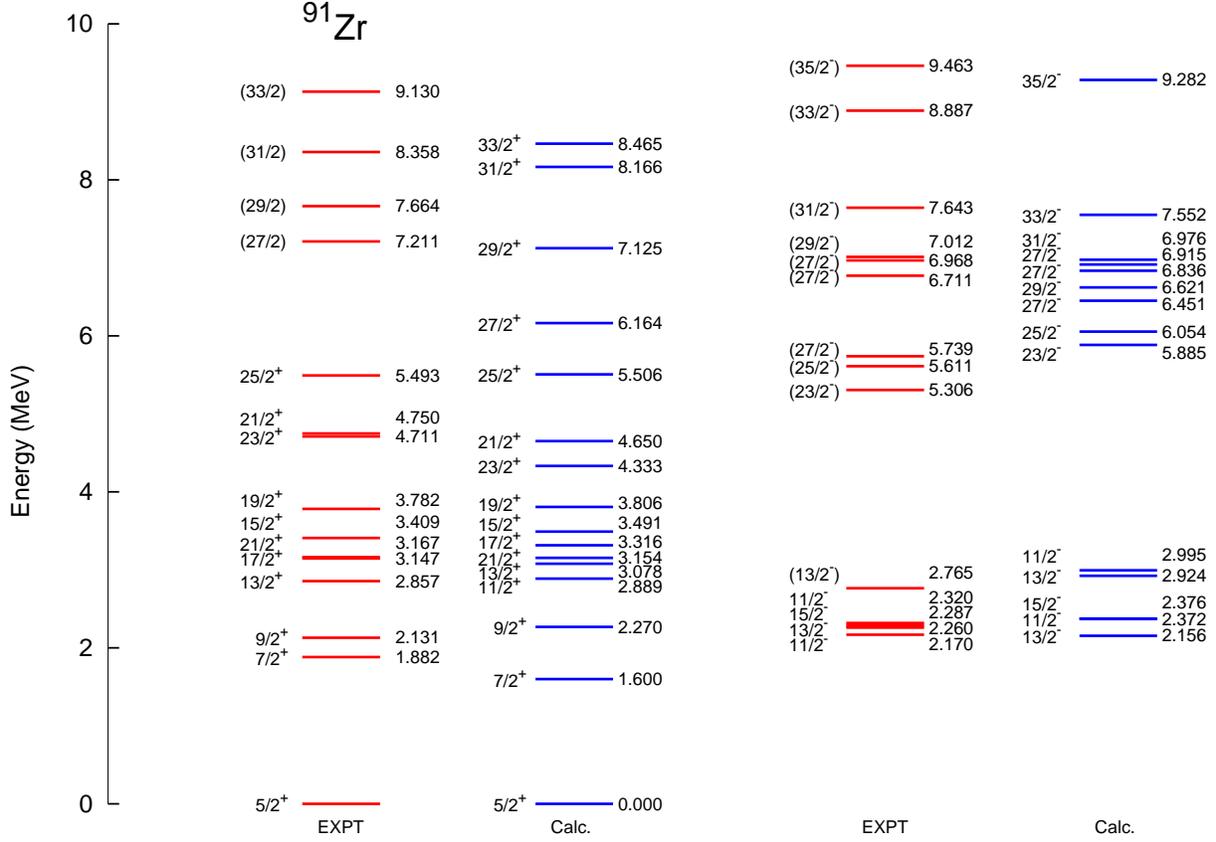}
\caption{ Comparison of  shell-model results for positive and negative parity states
with experimental data for $^{91}$Zr.
}
%\label{calculs_SM}      
\end{figure}

\begin{figure}
\begin{center}
\includegraphics[width=17.8cm]{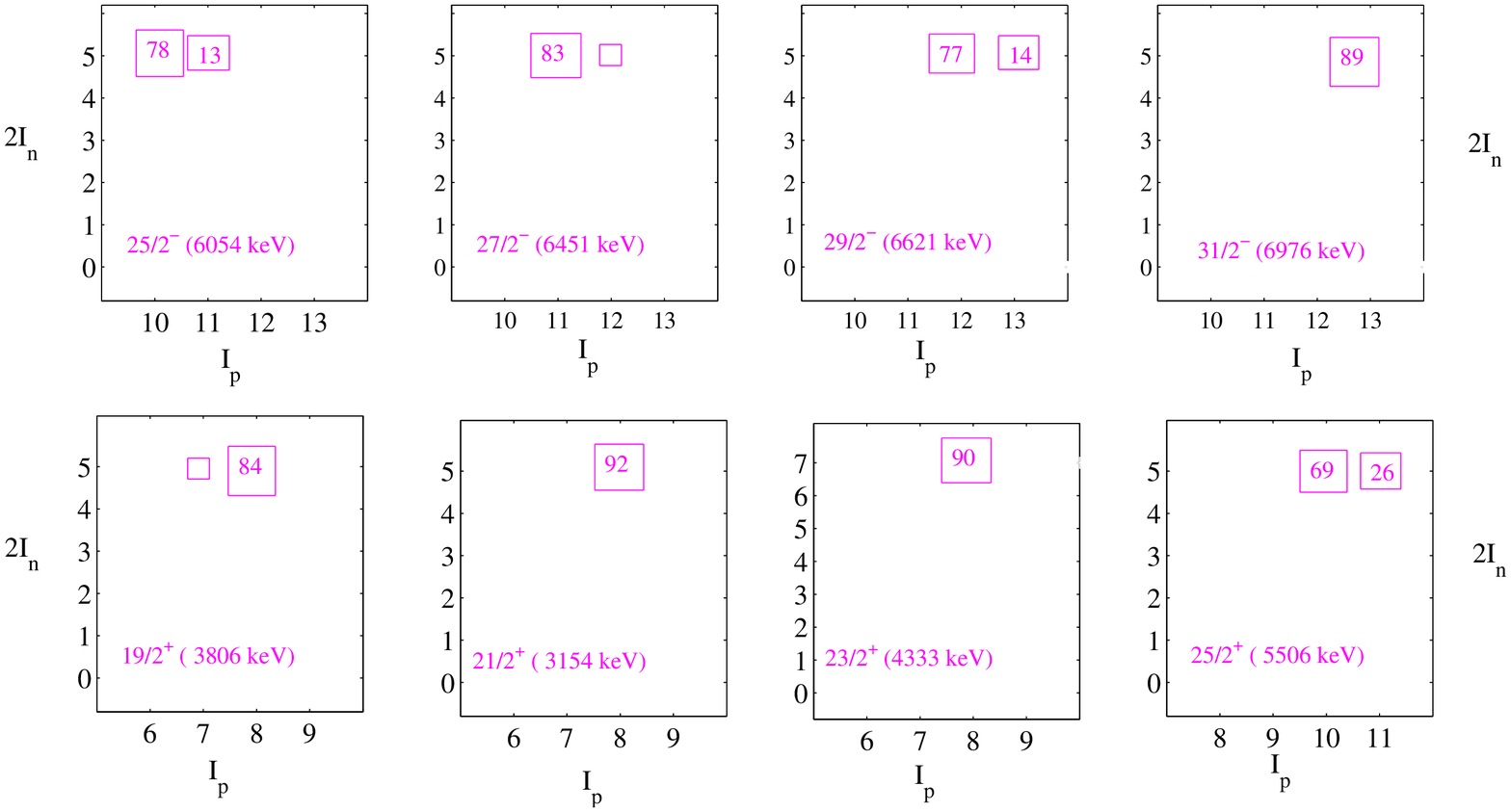}
\caption{ Decomposition of the total angular momentum of selected states of $^{91}$Zr into their $I_n \otimes I_p$ components. The percentage above 10\%
 are written inside the squares, drawn with an area proportional to it. Percentage below 5\% are not written.
}
\end{center}
%\label{calculs_SM}      
\end{figure}
%%%%%%%%%%%%%%%%%%%%%%%%%%%%%%%%%%%%%%%%%%%%%%%%%%%%%%%%%%
\section{Results and discussions}\label{discuss_SM}

\subsection{\bf Structure of $^{91}$Zr:}\label{discuss_SM1}

Results for energy levels of $^{91}$Zr are shown in Fig. 1. The energy levels of positive parity states $5/2^+$ - $33/2^+$,
are compared with the available experimental data. The overall agreement with experimental data is good.
The calculated $7/2_1^+$ is 282 keV, while $9/2_1^+$ is 139 keV above the experimental data. 
In the case of negative parity, the calculation predicts a flip of levels for $11/2_1^-$ and $13/2_1^-$. 
The experimentally observed plausible  state ($13/2_1^-$) at 2765 keV was predicted at 2924 keV. The calculated $35/2_1^-$ state 
is 181 keV lower than the experimental value.
 The ground state $5/2_1^+$ of $^{91}$Zr is showing single-particle character with $\nu d_{5/2}$ configuration.
The $7/2_1^+$ state has $\nu g_{7/2}$ configuration. The states $9/2_1^+$--$11/2_1^+$--$13/2_1^+$--$15/2_1^+$--$17/2_1^+$--$19/2_1^+$--$21/2_1^+$
mainly come from 
$\pi(f_{5/2}^{6}p_{3/2}^{4}g_{9/2}^{2})\otimes\nu (d_{5/2}^{1}$) configuration.
The low-lying isomeric $21/2_1^+$ state is formed with the fully aligned spin of two $g_{9/2}$ protons and one $d_{5/2}$ neutron.
The positive-parity states beyond the $21/2_1^+$ isomer can be understood with different configuration as the 
proton excitations across the $Z=38$ subshell and one neutron excitation from 
$\nu d_{5/2}$  to $\nu g_{7/2}$ orbital.
The structure of $25/2_1^+$ and $27/2_1^+$ states
 come from $\pi(f_{5/2}^{5}p_{3/2}^{4}p_{1/2}^{1}g_{9/2}^{2})\otimes\nu (d_{5/2}^{1})$ configuration.
The  $29/2_1^+$ state has  $\pi(f_{5/2}^{5}p_{3/2}^{4}p_{1/2}^{1}g_{9/2}^{2})\otimes\nu (g_{7/2}^{1})$ configuration.
From $29/2_1^+$ to $31/2_1^+$, proton empties from $p_{1/2}$ to $g_{9/2}$ and configuration for  $31/2_1^+$ becomes 
$\pi(f_{5/2}^{4}p_{3/2}^{4}g_{9/2}^{4})\otimes\nu (g_{7/2}^{1})$.
The states, $27/2_1^--29/2_1^--31/2_1^-$ come from $\pi(f_{5/2}^{5}p_{3/2}^{4}g_{9/2}^{3})\otimes\nu (d_{5/2}^{1})$,
while $33/2_1^-$ has $\pi(f_{5/2}^{5}p_{3/2}^{4}g_{9/2}^{3})\otimes\nu (g_{7/2}^{1})$ configuration. 
The $27/2_2^-$ state at 6711 keV has $\pi(f_{5/2}^{5}p_{3/2}^{4}g_{9/2}^{3})\otimes\nu (g_{7/2}^{1})$ configuration.

The analysis  of  wave functions also helps us to identify which type either proton or neutron pairs are broken to obtain the total angular
momentum of the calculated states.
The wave functions are decomposed in terms of the proton and neutron angular momenta $I_{p}$ and $I_{n}$. These
components are coupled to give the total angular momentum of each state. In the Fig. 2, we have shown results of positive  
and negative parity states of $^{91}$Zr. The positive parity states $19/2_1^+$ - $21/2_1^+$
come from $I_{p}= 8^+$ $\otimes$ $I_{n}= 5/2^+$. The dominant
components are 84\% ($19/2_1^+$), 92\% ($21/2_1^+$). The $23/2_1^+$ comes from $I_{p}= 8^+$ $\otimes$ $I_{n}= 7/2^+$ (90\%), while for $25/2^+$, the configuration is
$I_{p}= 10^+$ $\otimes$ $I_{n}= 5/2^+$ (69\%) and $I_{p}= 11^+$ $\otimes$ $I_{n}= 5/2^+$ (26\%). 
Decomposition of the total angular momenta for the negative parity states $25/2_1^-$ - $31/2_1^-$ are also shown in Fig. 2.
The $31/2_1^-$ state comes from $I_{p}= 13^-$ $\otimes$ $I_{n}= 5/2^+$ (89\%).
The experimental high-spin states (27/2), (29/2), (31/2) and (33/2)  at 7211.4, 7663.7, 8358.2, 9129.8 keV, respectively 
are plotted in the Fig. 1. Although it is not possible to confirm from the shell model it is positive or negative parity states, 
nevertheless we have plotted corresponding positive parity states 
predicted by shell model at 6164, 7125, 8166, 8465 keV, respectively. The configuration of $27/2_1^+$ is $I_{p}= 11^+$ $\otimes$ $I_{n}= 5/2^+$ (94.4\%);
$29/2_1^+$ is $I_{p}= 11^+$ $\otimes$ $I_{n}= 7/2^+$ (94.3\%); $31/2_1^+$ is $I_{p}= 12^+$ $\otimes$ $I_{n}= 7/2^+$ (79\%);
$33/2_1^+$ is $I_{p}= 8^+$ $\otimes$ $I_{n}= 17/2^+$ (41\%). The configuration for $33/2_1^+$ is
$\pi(f_{5/2}^{6}p_{3/2}^{4}g_{9/2}^{2})\otimes\nu (g_{9/2}^{-1}g_{7/2}^{1}d_{5/2}^{1}$).  The $15/2_1^-$ at 2287.8 keV has an isomeric states, the shell model calculation predicting it at 2376 keV, the configuration of this state
is  $\pi(f_{5/2}^{6}p_{3/2}^{4}p_{1/2}^{1}g_{9/2}^{1})\otimes\nu (d_{5/2}^{1})$. This configuration is responsible for predicting $11/2_1^-$--$13/2_1^-$--$15/2_1^-$ states. The next negative parity state
$17/2_1^-$ state is coming from $\pi(f_{5/2}^{5}p_{3/2}^{4}p_{1/2}^{2}g_{9/2}^{1})\otimes\nu (d_{5/2}^{1})$ configuration.

%%%In the table I, we have shown all energy levels as predicted by the shell model. While in the Figs.1-2, only levels corresponding to experimental
%%%data are plotted.

\begin{figure}
\includegraphics[width=17.0cm]{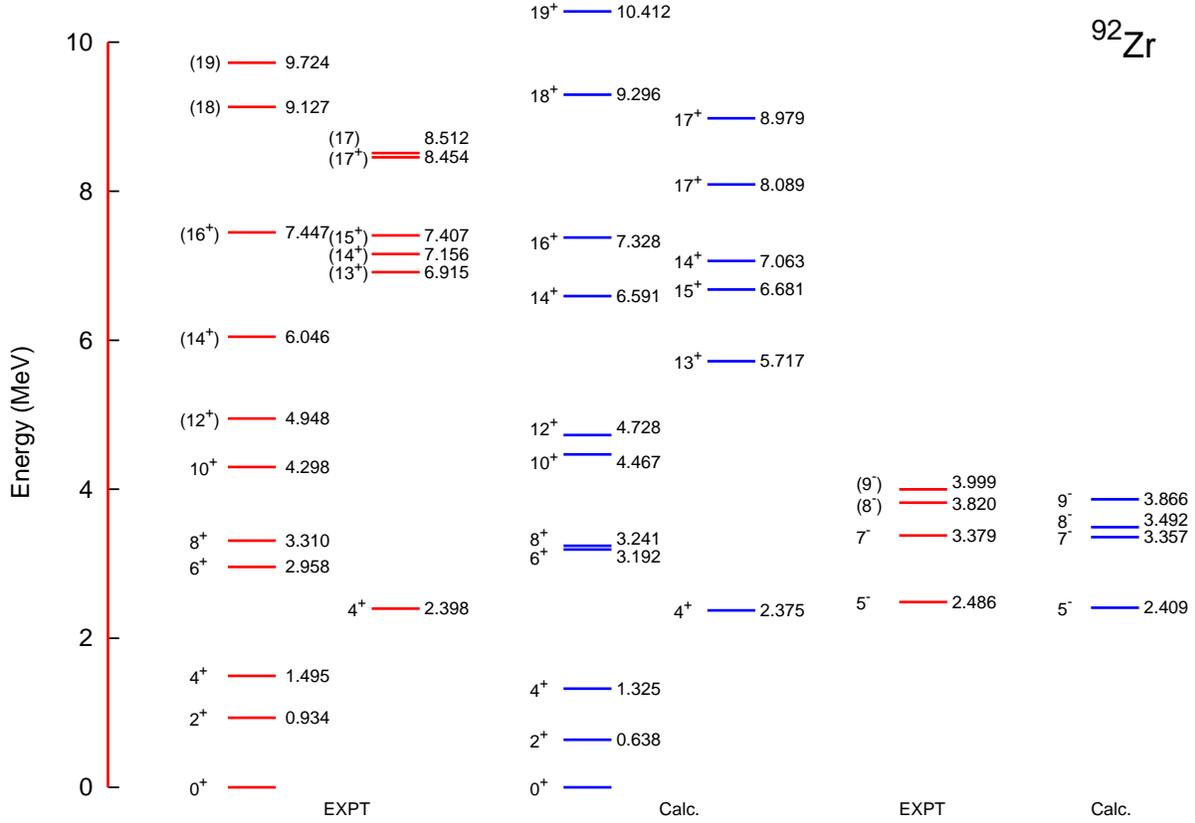}
\caption{ Comparison of  shell-model results for positive and negative parity states 
with experimental data for $^{92}$Zr.
}
\label{calculs_SM}      
\end{figure}
%%%%%%%%%%%%%%%%%%%%%%%%%%%%%%%%%%%%%%%%%%%%%%%%%%%%%%%%%%

\begin{figure}
\includegraphics[width=17.2cm]{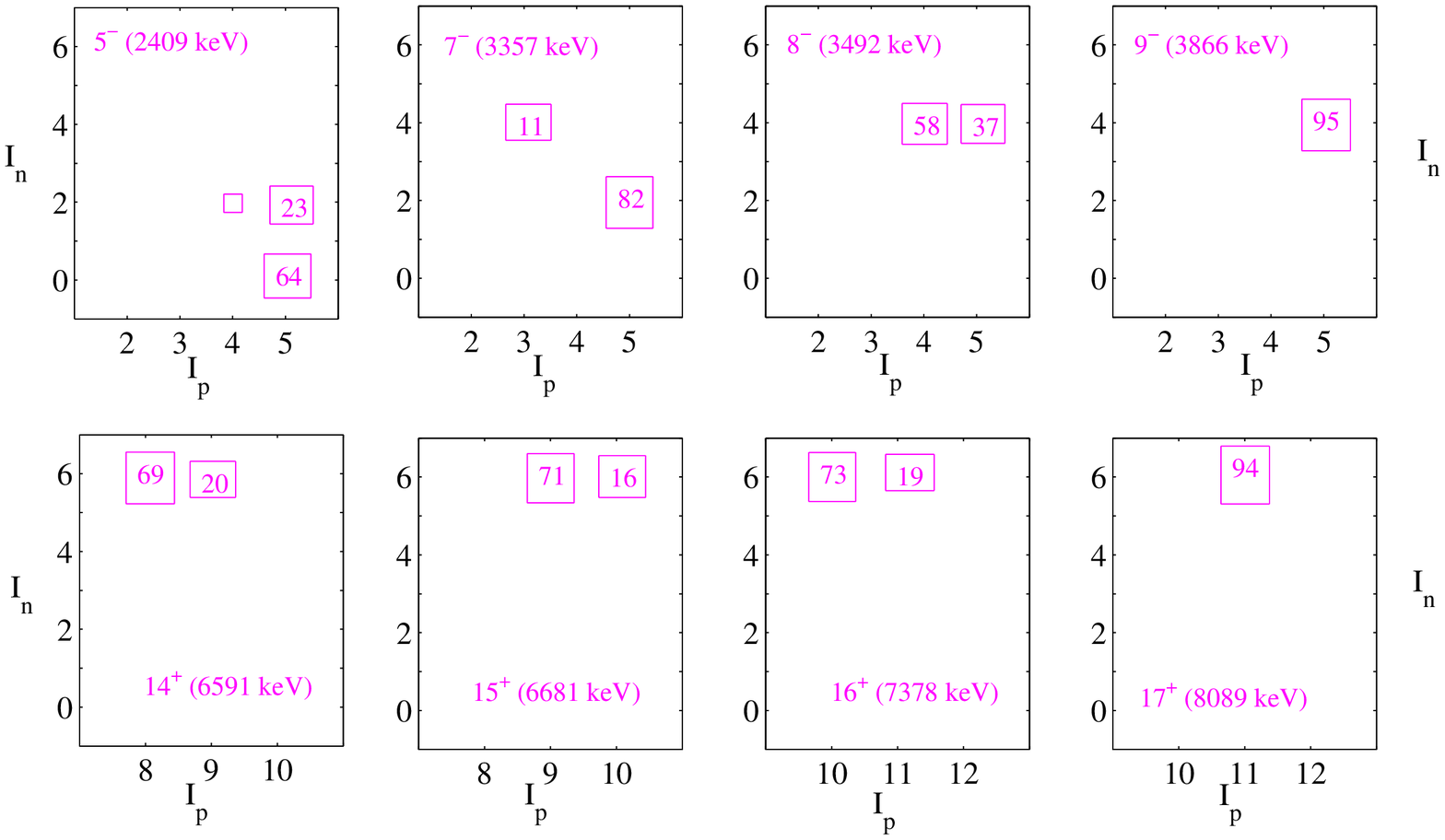}
\caption{ Decomposition of the total angular momentum of selected states of $^{92}$Zr into their $I_n \otimes I_p$ components. The percentage above 10\%
 are written inside the squares, drawn with an area proportional to it. Percentage below 5\% are not written.
}
%\label{calculs_SM}      
\end{figure}

\begin{figure}[h]
%\begin{sidewaysfigure}
%\begin{center}
\includegraphics[width=12.5cm,height=12.6cm,keepaspectratio]{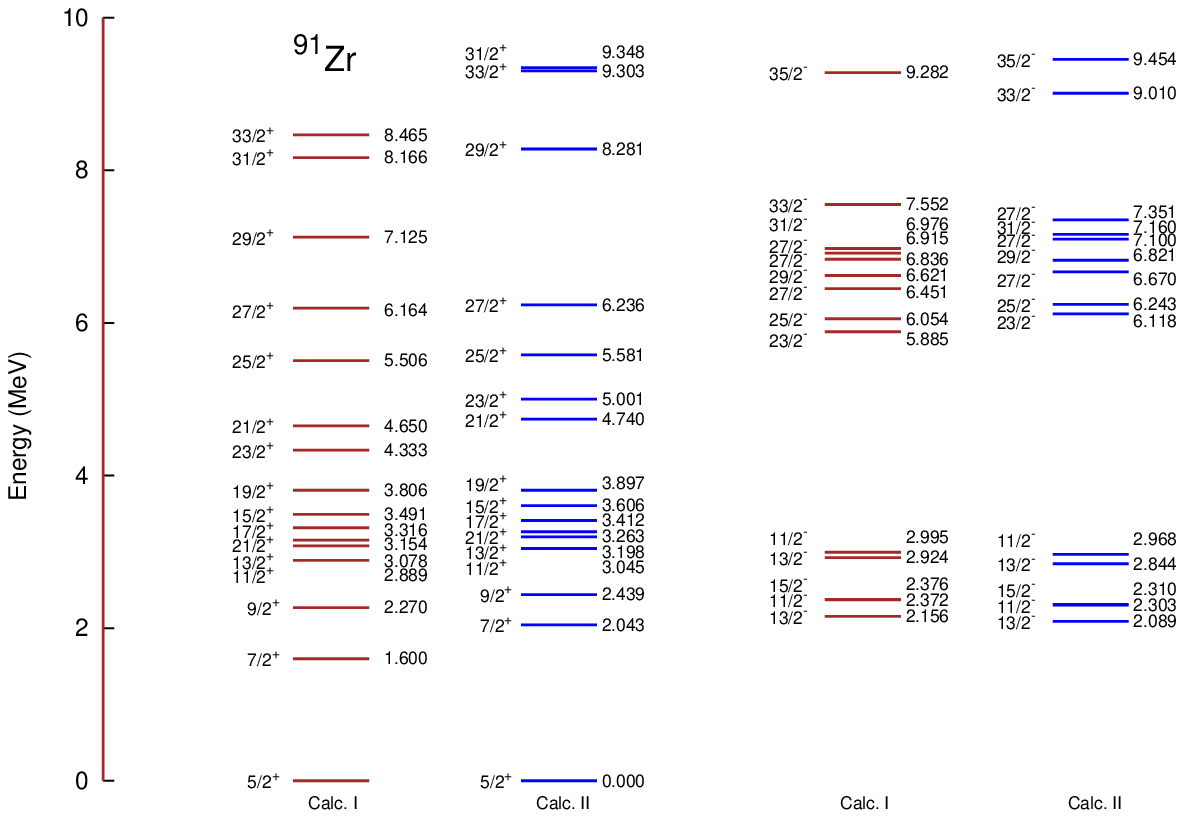}
\hspace{0.5mm}
\includegraphics[width=12.5cm,height=12.6cm,keepaspectratio]{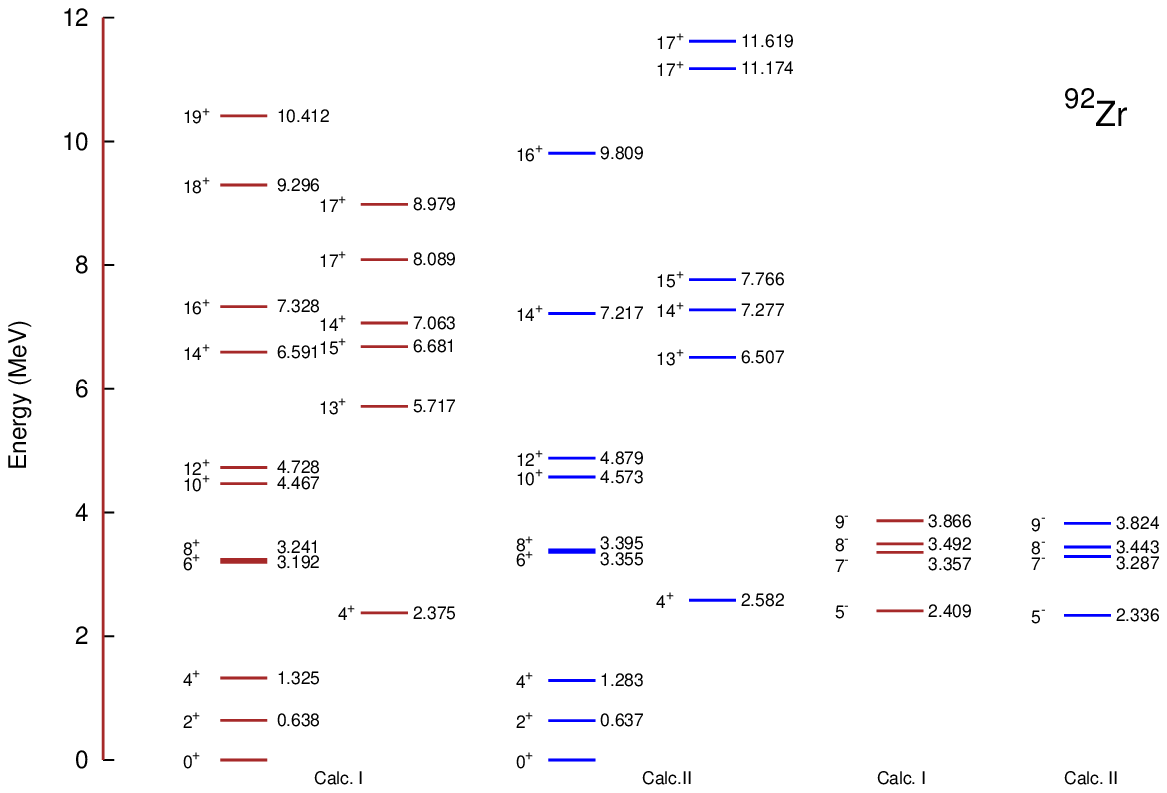}
%\end{center}
\caption{ Comparison of two different shell-model results for positive and negative parity states
for $^{91}$Zr and $^{92}$Zr. First calculation with one neutron excitation from
$g_{9/2}$ orbital to $d_{5/2}$ and $g_{7/2}$ orbitals (Calc. I) and second calculation with completely filled
$\nu g{9/2}$ orbital (Calc. II).}
%\end{sidewaysfigure}
\end{figure}

\subsection{\bf Structure of $^{92}$Zr:}\label{discuss_SM2}

The results of $^{92}$Zr are shown in Fig. 3.  The dominant configuration of $0_1^+$, $2_1^+$, and $4_1^+$ states
is $\nu (d_{5/2}^2)$. The $6_1^+$ and $8_1^+$ states are at a large gap of 1.5 MeV from $4_1^+$ state because
we need more energy to excite protons from $p_{1/2}^2$ to $g_{9/2}^2$ orbital to form these two states
([$\pi g_{9/2}^2]_{6^+,8^+}$). The higher angular momentum states $10_1^+$ and $12_1^+$  are associated with 
$\pi g_{9/2}^2 \otimes \nu d_{5/2}^2$  configuration. Beyond $12_1^+$, other high-spin states arise from excitation
of one proton/neutron across the $Z=38$/$N=56$ shell, i.e. one proton excitation from $(f_{5/2}p_{1/2})$
$\rightarrow$ $g_{9/2}$ and one neutron from $d_{5/2}$ $\rightarrow$ $g_{7/2}$.
The shell model calculations predict first three $14^+$ states at 4618, 6591, and 7063 -keV, respectively.
The $14^+$ state at 6591 keV is due to $\pi(f_{5/2}^{6}p_{3/2}^{4}g_{9/2}^2)$ $\otimes$ $\nu (d_{5/2}^{1}g_{7/2}^{1})$ configuration.
After looking at the structure and occupancy of  three eigen values of calculated
$14^+$ states. We found that $12^+$ (4728 keV) and  $14^+$ (6591 keV) have the same occupancy of $g_{7/2}$ and $d_{5/2}$ orbitals
[also $B(E2)$ value for this transition is large].  Thus we have taken
$14^+$ state at 6591 keV and $14^+$ state at 7063 keV in the Fig. 3 for the comparison with the experimental data. 
 The structure of $14^+$ state at 7063 keV is changed from $14^+$ state at 6591 keV
and it becomes  $\pi(f_{5/2}^{5}p_{3/2}^{4}p_{1/2}^{1}g_{9/2}^2)$  $\otimes$ $\nu (d_{5/2}^{1}g_{7/2}^{1})$.
The average occupancy of $d_{5/2}$ orbital for $12^+$ ( 5314 keV) and $14^+$ (4618 keV) states is $\sim$ 1.09, while
for $14^+$ (7063 keV) state it is 1.83. This is also reflected from the structure change from $14^+$ at 6591 keV to $14^+$ at 7063 keV.
The present study predicts that the $15_1^+$ and $17_1^+$ 
are fully neutron aligned states of the $\pi(f_{5/2}^{5}p_{3/2}^{4}p_{1/2}^{1}g_{9/2}^2)$ $\otimes$ $\nu (d_{5/2}^{1}g_{7/2}^{1})$ 
configuration. These states are built on $I_n$=$6^+$ [$\nu (d_{5/2}^{1}g_{7/2}^{1}$)].
The negative parity state $5_1^-$ comes from $\pi(f_{5/2}^{6}p_{3/2}^{4}p_{1/2}^{1}g_{9/2}^{1})$ configuration and the 
 $7_1^-$ to $9_1^-$ states come from $\pi(f_{5/2}^{6}p_{3/2}^{4}p_{1/2}^{1}g_{9/2}^{1})$  $\otimes$ $\nu (d_{5/2}^2)$. The high-spin
isomer $17_1^-$ observed in $^{92}$Zr is reported at 8.041 MeV  \cite{refiso}; we have calculated this state at
7.781 MeV. The main component of the wave function of this state is
$\pi(f_{5/2}^{5}p_{3/2}^{4}g_{9/2}^{3})\otimes\nu (d_{5/2}^{1}g_{7/2}^{1})$.

The wave function analysis for $^{92}$Zr is shown in Fig. 4. The $14^+$ (6591 keV) comes from 
$I_{p}= 8^+$ $\otimes$ $I_{n}= 6^+$ (87\%);  the $14^+$ (6591 keV) comes from 
$I_{p}= 8^+$ $\otimes$ $I_{n}= 6^+$ (69\%);
$15^+$  comes from  
$I_{p}= 9^+$ - $10^+$ $\otimes$ $I_{n}= 6^+$; $16_1^+$ comes from 
$I_{p}= 10^+$ - $11^+$ $\otimes$ $I_{n}= 6^+$ and $17_1^+$ comes from 
$I_{p}= 11^+$ $\otimes$ $I_{n}= 6^+$ (94\%).  The composition of negative parity state $5_1^-$ 
comes from  $I_{p}= 5^-$ ( 64\%) configuration. The $7_1^-$ comes from, 
$I_{p}= 5^-$ $\otimes$ $I_{n}= 2^+$ (82\%) and $I_{p}= 3_1^-$ $\otimes$ $I_{n}= 4^+$ (11\%).
The $8^-$ comes from, 
$I_{p}= 4^- - 5^-$ $\otimes$ $I_{n}= 4^+$ (82\%), while $9_1^-$ comes from 
$I_{p}= 5^-$ $\otimes$ $I_{n}= 4^+$ (95\%). 
 The experimentally observed high-spin plausible positive parity states (18) and (19)  at 9129.6, 9724.2 keV are
predicted by shell model at 9296,  10412 keV, respectively. The configuration of $18_1^+$ is $I_{p}= 12^+$ $\otimes$ $I_{n}= 6^+$ (77\%).
For $19_1^+$ ,the major configuration is $I_{p}= 14^+$ $\otimes$ $I_{n}= 6^+$ (52\%) and other dominant configuration is
$I_{p}= 13^+$ $\otimes$ $I_{n}= 6^+$ (35\%).

 There are two-ways to generate high-spin angular momentum due to single-particle effects. First is by pair-breaking and second is due to particle-hole
 excitations. Thus creating  one-hole ($g_{9/2}^{-1}$) configuration is important for generating high-spin states.

 The calculated energy levels of low-lying states for $^{92}$Zr are more compressed. Thus, we have also
performed second calculation (Calc. II) for $^{91}$Zr and $^{92}$Zr without neutron excitation from $\nu g_{9/2}$ to $\nu d_{5/2}$ orbital,
i.e. we have completely filled the $\nu g{9/2}$ orbital.
The results are shown in Fig. 5. In the case of $^{92}$Zr, the results for low-lying states
$6_1^+$ -$8_1^+$-$10_1^+$-$12_1^+$ show good agreement with the experimental data.
Previously, shell model results for $^{92}$Zr with the same effective interaction (TBMEs) for the
proton ($p_{1/2}$, $g_{9/2}$) and neutron ($d_{5/2}$, $s_{1/2}$) orbitals were reported in Ref. \cite{int3}. The low-lying states
in this restricted model space show good agreement with the experimental data,
although, high-spin states are not in good agreement without neutron excitation from $\nu g_{9/2}$ to $\nu d_{5/2}$ orbital,
as we have shown in Figs. 1 and 3. 
The second set of calculation predicting $6_1^+$ and $6_2^+$  at 3355 and 3944 keV, respectively. Thus
reported  $6^+$ state is the first state.

\subsection{\bf Electromagnetic properties }\label{discuss_SM4}

To test further  the quality of our SM wave functions, we
have calculated several $E2$ transitions, which are listed
in Table I. First, we have performed the calculations with the
standard effective charges, i.e., $e_{\pi} =1.5e$ and $e_{\nu} =0.5e$.
Since the calculation without truncation is not feasible in the present model space, thus to account for
the missing mixing of configurations, we have also performed calculations with the increased effective charges
$e_{\pi} =1.8e$ and $e_{\nu} =0.8e$.
This might be due to missing neutron orbitals below 
$p_{1/2}$, thus effect of core polarization is important. It also gives explanation to  adjust
the effective charges such that the ground state transition strength is correctly reproduced as in the experimental data \cite{ref1,ref2,wernerplb}.  
In Table I, we have listed both results. 
These results indicating deviation with experimental data even with the enhanced values of effective charges.

The calculated $B(E2;8^{+}_{1} \rightarrow 6^{+}_{1}$) values with two set of effective charges are small in comparison to the experimental
data.
In Table II, we have listed quadrupole and magnetic moment for the ground state and few excited states. 
The results are in good agreement with the experimental data for $^{91}$Zr. 
From Table II, it is also possible to compare recently available experimental data for the  magnetic moment with the calculated values. 
The experimental value  for $\mu(2_1^{+})$ is -0.360(20), while corresponding calculated value is -0.458 ($g_s^{eff}= 0.7 g_s^{free}$).
For $2_2^+$ state, the experimental value  for $\mu (2_2^{+})$ is +1.5(10), while corresponding calculated value is +0.972 ($g_s^{eff}= 0.7 g_s^{free}$).
Similarly for  $4_1^+$ state the experimental value  for $\mu(4_1^{+})$ is -2.0(4), while corresponding calculated value is -1.456 ($g_s^{eff}= 0.7 g_s^{free})$.

\begin{table}[hbtp]
\caption{Experimental \cite{a=92} and calculated $B(E2)$ values in W.u. with different set of effective charges and B(M1) with $g_s^{eff}=g_s^{free}$ {Set. I}/ $g_s^{eff}= 0.7 g_s^{free}$ {Set. II}.}
\label{t_be2}
\begin{center}
\begin{tabular}{rccccc}
\hline
Nucleus  & ~~Transition & ~~Expt. &~~ Set. I  &~~ Set. II &\\   
    & ~~  & ~~ &$e_{\pi} =1.5e$,   & $e_{\pi} =1.8e$ &\\ 
    & ~~  & ~~ &$e_{\nu} =0.5e$   & $e_{\pi} =0.8e$ &\\ 
\hline
$^{91}$Zr &$B(E2;7/2^{+}_{1} \rightarrow 5/2^{+}_{1}$)\hspace{1.cm} & 7.5 (13) & 1.44 & 2.34 \\ 
           &$B(E2;9/2^{+}_{1} \rightarrow 5/2^{+}_{1}$)\hspace{1.cm} & 4.2 (6) & 3.32 & 4.90\\ 
           &$B(E2;13/2^{+}_{1} \rightarrow 9/2^{+}_{1}$)\hspace{1.cm} & $>$ 0.0079  & 6.31 & 9.51 \\ 
          &$B(E2;21/2^{+}_{1} \rightarrow 17/2^{+}_{1}$)\hspace{1.cm} & 4.3 (7) & 1.72 & 2.58 \\ 
\hline
$^{92}$Zr &$B(E2;2^{+}_{1} \rightarrow 0^{+}_{1}$)\hspace{1.cm} & 6.4(6) & 4.67 &  7.94 \\ 
            &$B(E2;4^{+}_{1} \rightarrow 2^{+}_{1}$)\hspace{1.cm} & 4.05(12) & 1.97  &  3.69 \\ 
             &$B(E2;8^{+}_{1} \rightarrow 6^{+}_{1}$)\hspace{1.cm} & 3.59(22) & 1.18 & 1.85  \\
              &$B(E2;12^{+}_{1} \rightarrow 10^{+}_{1}$)\hspace{1.cm} & $\geq$ 0.056 & 2.56  & 4.49 \\
              &$B(E2;14^{+}_{2} \rightarrow 12^{+}_{1}$)\hspace{1.cm} &  &  0.029 & 0.047 \\
               &$B(M1;2^{+}_{2} \rightarrow 2^{+}_{1}$)\hspace{1.cm} & 0.46(15) & 0.164   & 0.117 \\
\hline
           
\end{tabular}
%\label{be2}
\end{center}
\end{table}

\begin{table}[hbtp]
\caption{Comparison of the calculated and the experimental \cite{a=92}  quadrupole ($eb$) and magnetic ($\mu_N^2$) moments. The calculated quadrupole moment
with $e_{\pi} =1.5e$, $e_{\nu} =0.5e$ {Set. I} / $e_{\pi} =1.8e$, $e_{\nu} =0.8e$ {Set. II} and magnetic moment with $g_s^{eff}=g_s^{free}$ {Set. I}/ $g_s^{eff}= 0.7 g_s^{free}$ {Set. II}.}
\label{t_be2}
\begin{center}
\begin{tabular}{cccccccc}
\hline
& $J^{\pi}$   & $Q_{s,exp}$ & $Q_{s,Set.I} $ & $Q_{s,Set.II} $ & $\mu_{exp.}$ & $\mu_{Set. I}  $  & $\mu_{Set. II}$  \\   
\hline
$^{91}$Zr & $5/2_1^{+}$   &-0.176(3)  & -0.203    & -0.273   &  -1.30362(2) & -1.310  & -0.900\\ 
$^{92}$Zr & $2_1^{+}$      & N/A      & +0.158    & +0.209   &  -0.360(20)  & -0.745  & -0.458 \\ 
& $2_2^{+}$                & N/A      & +0.248    & +0.319   &  +1.5(10)    & +0.962  & +0.972  \\ 
& $4_1^{+}$                 & N/A     & -0.095    & -0.136   &  -2.0(4)     & -2.122  & -1.456 \\ 
& $4_2^{+}$                  & N/A    & +0.359    & +0.448   &  N/A         &  +2.614 & +2.498\\            
\hline
           
\end{tabular}
%\label{be2}
\end{center}
\end{table}

\section{Conclusions}\label{fin}

Structure of high spin levels  of  $^{91,92}$Zr have been studied with large-basis shell model calculations 
for neutrons excitation across $N=50$ shell. 
Following broad conclusions can be drawn from the present work.
\begin{itemize}

\item { The present study reveals the importance of inclusion of $d_{5/2}$ and $g_{7/2}$ orbitals in the model space for high-spin states
beyond $\sim$ 4.5 MeV. }

\item { The present shell model results are in good agreement with the proposed configurations in Ref. \cite{9192Zr}.
The structure of $25/2_1^+$ in $^{91}$Zr is mainly from $\pi(f_{5/2}^{5}p_{3/2}^{4}p_{1/2}^{1}g_{9/2}^{2})\otimes\nu d_{5/2}^{1}$ configuration.
The structure of $27/2_1^- -27/2_1^-- 31/2_1^-$ states come from $\pi(f_{5/2}^{5}p_{3/2}^{4}g_{9/2}^{3})\otimes\nu d_{5/2}^{1}$ configuration,
while $33/2_1^-$
have $\pi(f_{5/2}^{5}p_{3/2}^{4}g_{9/2}^{3})\otimes\nu (g_{7/2}^{1}$) configuration.}

\item { For $^{92}$Zr, 
$15_1^+$ to $17_1^+$ states
are fully aligned state of the $\pi(f_{5/2}^{5}p_{3/2}^{4}p_{1/2}^{1}g_{9/2}^2)$ $\otimes$ $\nu (d_{5/2}^{1}g_{7/2}^{1})$ configuration.}

\item {We have calculated electromagnetic properties between different transitions for few high-spin states, wherever
experimental data is available. These results indicating deviation with experimental data  even with the enhanced values of effective charges.}

\end{itemize}

\section*{Acknowledgment}
I would like to acknowledge financial support from faculty initiation grant.
Thanks are due to Y.H. Zhang and Y. H. Qiang  for useful suggestions during this work.
I would like to thank Mirshod Ermamatov  and V.K.B. Kota for a careful reading of the manuscript.\\

%\section*{References:}

%\end{comment}
%\appendix

\begin{thebibliography}{99}

%\bibitem{ref0} N. Yoshinaga, K. Higashiyama, and P. H. Regan, Phys. Rev. C {\bf 78}, 044320 (2008).
\bibitem{ref1} J. D. Holt {\it et al.}, Phys. Rev. C {\bf 76}, 034325 (2007).
\bibitem{ref2} J. N. Orce {\it et al.}, Phys. Rev. Lett. {\bf 97}, 062504 (2006).
\bibitem{ref3} C. Fransen {\it et al.}, Phys. Rev. C {\bf 71}, 054304 (2005).
\bibitem{ref4} J. N. Orce  {\it et al.}, Phys. Rev. C {\bf 82}, 044317 (2010).
\bibitem{ref5} P.H. Regan {\it et al.}, AIP Conference Proceedings {\bf 819}, 35 (2006).
\bibitem{ref6} A. D. Ayangeakaa, Master thesis "Gamma-Ray spectroscopy of shell model states in $^{91,92}$Zr",
University of Surrey, (2006).

\bibitem{9192Zr} Z.G. Wang {\it et al.}, Phys. Rev. C {\bf 89}, 044308 (2014).
                
\bibitem{honma} M. Honma, T. Otsuka, T. Mizusaki and M. Hjorth-Jensen, Phys. Rev. C {\bf 80}, 064323 (2009).
          
\bibitem{yoshi} N. Yoshinaga, K. Higashiyama, and P.H. Regan, Phys. Rev. C {\bf 78}, 044320 (2008).
                                
\bibitem{nakada} H. Nakada, Prog. Theor. Phys. (Kyoto), Suppl. {\bf 196}, 371 (2012).
\bibitem{gepcs} P.C. Srivastava,  J. Phys. G: Nuclear and Particle Physics {\bf 39} , 015102 (2012).
\bibitem{mpla} P.C. Srivastava, Mod. Phys. Lett. A {\bf 27},  1250061 (2012).
\bibitem{62ga} P.C. Srivastava, R. Sahu and V.K.B. Kota,  Eur. Phys. J. A : Hadrons and Nuclei,  {\bf 51}, 3 (2015).
\bibitem{1a} E. Sahin {\it et al.}, Nucl. Phys. A {\bf 893}, 1 (2012).
              
\bibitem{88Zr} S. Saha {\it et al.}, Phys. Rev. C {\bf 86}, 034315 (2012).
                 
\bibitem{89Zr} S. Saha {\it et al.}, Phys. Rev. C {\bf 89}, 044315 (2014).
                
\bibitem{89Nb} P. Singh {\it et al.}, Phys. Rev. C {\bf 90}, 014306 (2014).
                
\bibitem{91Nb} P.W. Luo {\it et al.}, Phys. Rev. C {\bf 89}, 034318 (2014).
\bibitem{pittel} P. Federman and S. Pittel, Phys. Lett. B {\bf 69}, 385 (1977).
\bibitem{patt} N.S. Pattabiraman {\it et al.}, Phys. Rev. C {\bf 65}, 044324 (2002).
\bibitem{pajtler} M. Varga Pajtler {\it et al.}, Nucl. Phys. A {\bf 941}, 273 (2015).
\bibitem{bingham} C.R. Bingham and M.L. Halbert, Phys. Rev. C {\bf 2}, 2297 (1970).
\bibitem{pcappb} P.C. Srivastava, V. Kumar and M. J. Ermamatov, Acta. Phys. Pol. B {\bf 47}, 2151 (2016).
\bibitem{1b} M.-G. Porquet{\it et al.}, Phys. Rev. C {\bf 84}, 054305 (2011).
\bibitem{1c}E. A. Stefanova et al., Nucl. Phys. A {\bf 669}, 14 (2000).
\bibitem{2}E. A. Stefanova et al., Phys. Rev. C {\bf 62}, 054314 (2000).
\bibitem{3}J. Reif, G. Winter, R. Schwengner, H. Prade, and L. Käubler, Nucl. Phys. A {\bf 587}, 449 (1995).
\bibitem{4}E. K. Warburton, J. W. Olness, C. J. Lister, R. W. Zurmühle, and J. A. Becker, Phys. Rev. C {\bf 31}, 1184 (1985).
\bibitem{5}N. S. Pattabiraman et al., Phys. Rev. C {\bf 65}, 044324 (2002).
\bibitem{6}S. S. Ghugre and S. K. Datta, Phys. Rev. C {\bf 52}, 1881 (1995).
\bibitem{7}I. P. Johnstone and L. D. Skouras, Phys. Rev. C {\bf 55}, 1227 (1997).
\bibitem{8}H. A. Roth et al., Phys. Rev. C {\bf 50}, 1330 (1994).
\bibitem{9}K. Sieja, F. Nowacki, K. Langanke, and G. Mart\' inez-Pinedo, Phys. Rev. C {\bf 79}, 064310 (2009).
\bibitem{10} P. Federman and S. Pittel, Phys. Rev. C {\bf 20}, 820 (1979). 
\bibitem{bansal64} R.K. Bansal and J.B. French, Phys. Lett. {\bf 11}, 145 (1964).
\bibitem{otsu1} T.Otsuka {\it et al.}, Phys. Rev. Lett. {\bf 87}, 082502 (2001).
\bibitem{otsu2} T.Otsuka {\it et al.}, Phys. Rev. Lett. {\bf 95}, 232502 (2005).
\bibitem{prog3} T. Faestermann, M. G\'orska, H. Grawe, Prog. Part. Nucl. Phys. {\bf 69}, 85 (2013).
\bibitem{talmi} I. Talmi, Simple Models of Complex Nuclei, Harwood academic publishers, Switzerland (1993). 
\bibitem{wernerplb}V. Werner et al., Phys. Lett. B {\bf 550}, 140 (2002).
 \bibitem{9495Mo} Y.H. Zhang {\it et al.}, Phys. Rev. C {\bf 79}, 044316 (2009).
\bibitem{nushell} B.A. Brown, W.D.M. Rae, E. McDonald and M. Horoi, NuShellX@MSU.
 \bibitem{int1} A. Hosaka, K.-I. Kubo, and H. Toki, Nucl. Phys. A {\bf 444}, 76 (1985).
\bibitem{int2}  X. Ji and B. H. Wildenthal, Phys. Rev. C {\bf 37}, 1256 (1988).
\bibitem{int3} D. H. Gloeckner, Nucl. Phys. A {\bf 253}, 301 (1975).
\bibitem{int4} F. J. D. Serduke, R. D. Lawson, and D. H. Gloeckner, Nucl. Phys. A {\bf 256}, 45 (1976).
\bibitem{gwbxg_cite} D. Bucurescu {\it et al.}, Phys. Rev. C {\bf 76}, 064301 (2007).
\bibitem{refiso}  G. Korschinek {\it et al.}, Pro. of the Inter.
Conf. on Nuclear Structure, Tokyo, 1977.
\bibitem{a=92}  C.M. Baglin, Nucl. Data Sheets {\bf 113}, 2187 (2012).
%%\bibitem{nndc} http://www.nndc.bnl.gov/ensdf/
%%\bibitem{seweryniak} D. Seweryniak et al., Phys. Rev. Lett.  {\bf 99}, 022504 (2007).

\end{thebibliography}
\end{document}